# Performance of the VERITAS experiment

**Nahee Park**[*] **for the VERITAS Collaboration**[†]
*The University of Chicago*
*E-mail:* nahee@uchicago.edu

VERITAS is a ground-based gamma-ray instrument operating at the Fred Lawrence Whipple Observatory in southern Arizona. With an array of four imaging atmospheric Cherenkov telescopes (IACTs), VERITAS is designed to measure gamma rays with energies from $\sim$ 85 GeV up to $>$ 30 TeV. It has a sensitivity to detect a point source with a flux of 1% of the Crab Nebula flux within 25 hours. Since its first light observation in 2007, VERITAS has continued its successful mission for over seven years with two major upgrades: the relocation of telescope 1 in 2009 and a camera upgrade in 2012. We present the performance of VERITAS and how it has improved with these upgrades.

---

[*]Speaker.
[†]veritas.sao.arizona.edu

## 1. VERITAS operation and upgrades

VERITAS (the Very Energetic Radiation Imaging Telescope Array System) is an array of four imaging atmospheric Cherenkov telescopes (IACTs) located at the Fred Lawrence Whipple Observatory in southern Arizona ($30°40'N$ $110°57'W$, 1268 m a.s.l.) [1]. It is designed to study astrophysical sources of gamma-ray emission in the energy range from $\sim$ 85 GeV up to > 30 TeV by measuring the Cherenkov light generated by particle showers initiated by primary gamma rays in the atmosphere. Each of the four telescopes covers a field of view of 3.5° with a 499-pixel photomultiplier tube (PMT) camera at the focal plane, collecting light from a 12 meter diameter reflector consisting of segmented mirrors. A coincident Cherenkov signal triggered by at least two out of four telescopes is required to trigger an array-wide read-out of the PMT signals [2].

The array was commissioned in 2007 and has continued its successful mission for over seven years with continuous improvements in hardware, operation, calibration and analysis methods. There were two major hardware upgrades that changed the properties of the telescope array significantly: the relocation of telescope 1 in 2009 and a camera upgrade in 2012. The relocation of telescope 1 was motivated to make the array more symmetric, increasing the sensitivity by augmenting the stereo observation of the air showers [3]. The camera upgrade in 2012 replaced the PMTs with high quantum efficiency PMTs, which increased the Cherenkov photon collection efficiency by 50% [4].

This paper focuses on describing the performance of VERITAS, and how it changed with the two major upgrades. From here, we will refer to the period before the relocation of telescope 1 (2007/09/01-2009/08/31) as V4, the period after the relocation of telescope 1 and before the camera upgrade (2009/09/01-2012/08/31) as V5 and after the camera upgrade (2012/09/01-present) as V6.

## 2. Observation and analysis for performance study

VERITAS has carried out designated observations of the Crab Nebula in each hardware configuration for performance studies. The Crab Nebula is one of the brightest sources in the energy range of VERITAS. Although changes in the gamma-ray flux from the Crab Nebula have been observed in the GeV energy range by space-based instruments [5, 6], no significant flux changes were measured in the TeV energy range during major flare activity [7]. The brightness and steadiness of the Crab Nebula make it a good source to study the properties of the VERITAS telescopes in different observational conditions and in different array configurations. From 2007, VERITAS has taken over 69 hours of data on the Crab Nebula under various different observational conditions for monitoring the performance of the array. The combined data set of calibration data and scientific data taken on the Crab Nebula (total $\sim$ 200 hours) was used for this performance study.

The standard analysis of VERITAS data is described in [8]. VERITAS maintains two complete analysis packages, which allows independent cross checking of the results. The standard analysis uses box cuts with scaled parameters from a moment analysis [9] for event selections. Depending on the strength and expected spectral index of a source, different cuts are chosen *a priori*. While cuts can be optimized for specific scientific goals, the standard analysis packages of VERITAS [10] provide at least three sets of cuts (soft, medium and hard cuts) for general cases. Soft cuts are optimized for sources with a spectral index of −3.5 to −4.0, with reasonably low energy threshold values. Medium cuts are most widely used and should be near-optimal for the bulk of the sources of interest. They are optimized for sources of 2-10% of the Crab Nebula strength with an index of $-2.5$ to $-3.0$. Hard cuts are optimized to detect sources weaker than 2% of the Crab Nebula strength by selecting the purest gamma-ray samples with the highest energy threshold among the three sets of cuts. Medium and hard cuts are adjusted to have similar energy threshold values between epochs. These cuts will be used to describe the performance of the VERITAS array in this paper after consistency check with an independent analysis package.

## 3. Performance of the VERITAS experiment

### 3.1 Sensitivity of VERITAS

The differential sensitivity curves for the V6 epoch of VERITAS are shown in Figure 1a for three sets of cuts. The energy range was divided to have four bins per decade. Differential sensitivities are calculated with Crab Nebula data taken at high elevation to estimate the weakest source that can be detected at 5 $\sigma$ significance within 50 hours of observing time in each energy bin. Li & Ma's likelihood ratio method [11] was used to calculate the significance for each energy bin, provided that the bin contains at least 10 gamma-ray excess events in the source region. Events with small energy bias were selected for the calculation. Calculations with high elevation Crab data show that soft cuts provide the highest sensitivity at low energy (around a few hundred GeV) while hard cuts provide the best sensitivity at energies higher than 600 GeV.

A comparison of differential sensitivities between epochs demonstrates improvements in the sensitivities for all of the cuts with the VERITAS upgrades. The comparisons for medium cuts are shown in Figure 1b as an example. The plot shows that while the energy threshold values for each epoch are similar, the sensitivity for events with energies lower than 2 TeV improves dramatically. The improvement from V4 to V5 is mainly due to additional background rejection power gained by enhanced stereo imaging of the air shower. Improvement of the angular resolution, from the reduction of almost parallel images, contributed to the enhanced sensitivity as well. The improvement from V5 to V6 is from better morphological separation between gamma-ray and hadronic induced air showers due to the larger photon collection efficiency of the high quantum efficiency PMTs.

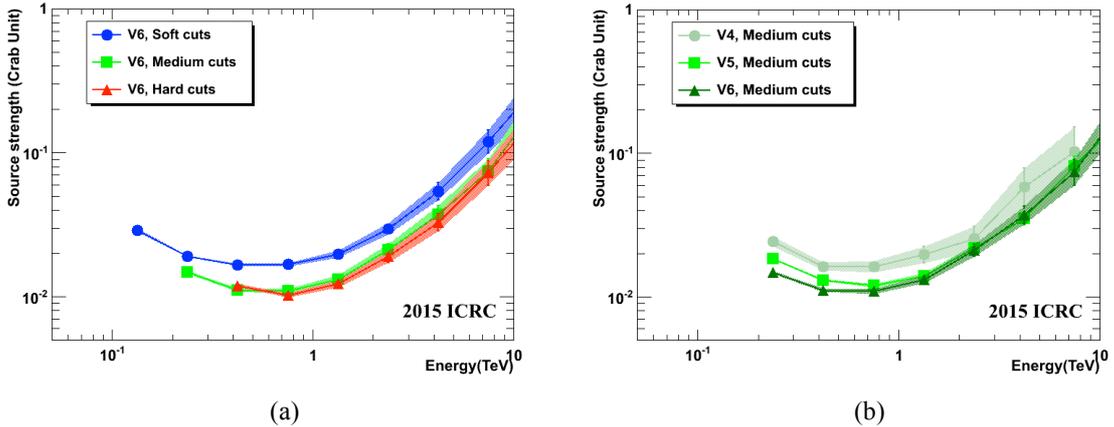

(a)        (b)

Figure 1: *(a):* Differential sensitivity estimated with standard cuts by using Crab Nebula data with elevation higher than 70°. We calculate the strength of the source as a percentage of the flux of the Crab Nebula in a given energy bin such that the source would be detected with 5 $\sigma$ significance after 50 hours of observing time. *(b):* Differential sensitivity with medium cuts for three different epochs.

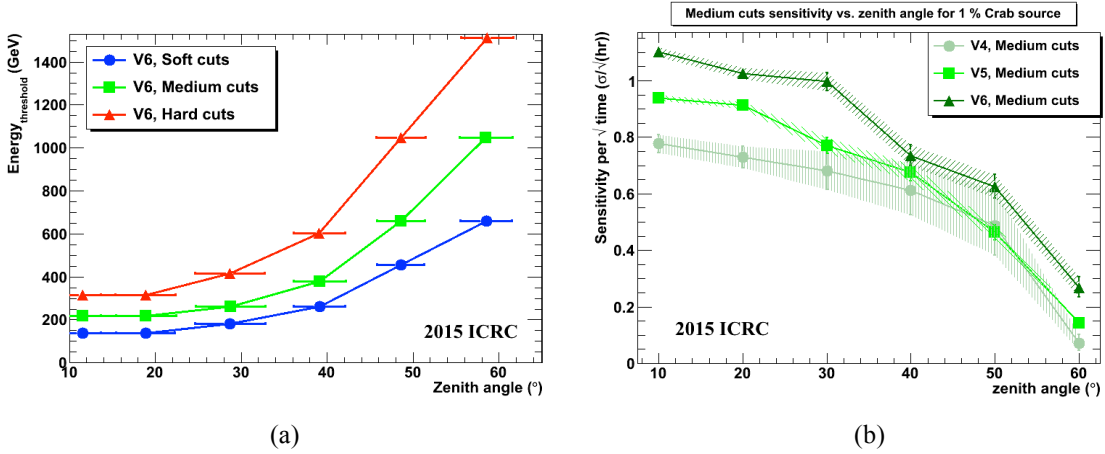

        (a)                                 (b)

Figure 2: *(a):* Energy threshold dependence on zenith angle for the V6 epoch. *(b):* Sensitivity dependence on zenith angle for medium cuts for the three epochs. The error bars are larger for the V4 epoch due to the small sample size.

The sensitivity of an IACT array changes with observing conditions such as the zenith and azimuth angle of the source, the offset of the source from the center of the camera, and the background light levels. Among these, the dependence of the sensitivity is largest for zenith angle and offset of the source. To study the performance of the array in these conditions, we observed the Crab Nebula at different zenith angles and offset angles.

Air showers initiated at larger zenith angles propagate through a thicker atmosphere and the Cherenkov light generated from these showers will be absorbed more. As a result, observations at large zenith angles have a higher energy threshold as shown in Figure 2a. The inclined showers also generate a larger pool of Cherenkov light, which increases the effective area. Understanding the zenith angle dependent changes of the performance is important for studying various sources culminating in different zenith angles, as well as for optimizing the observational strategies for specific scientific goals. For example, studies of high energy gamma-ray emission from a relatively bright source with a hard index can benefit from large zenith angle observations because of the large effective area despite the higher energy threshold.

Figure 2b shows the dependence of sensitivity on zenith angle for all epochs derived from Crab Nebula data. Here sensitivity is defined as a significance of gamma-ray emissions in square-root of exposure time ($\sigma/\sqrt{\text{hour}}$). Figure 2b shows the dependence of the sensitivity on the zenith angle for weak sources (1% of Crab Nebula strength) analyzed with medium cuts. The figure shows that the sensitivity is reduced as zenith angle increases. This is because large zenith angle observation has higher energy threshold values and worse directional reconstruction with the standard recon-struction method. Generally other reconstruction methods, such as the displacement method [12], are used for the analysis of large zenith angle data to improve the quality of shower reconstruction. The figure also shows that there are clear improvements of sensitivity as the configuration of VERITAS changed from V4 to V5 to V6 for all observations taken at different zenith angles.

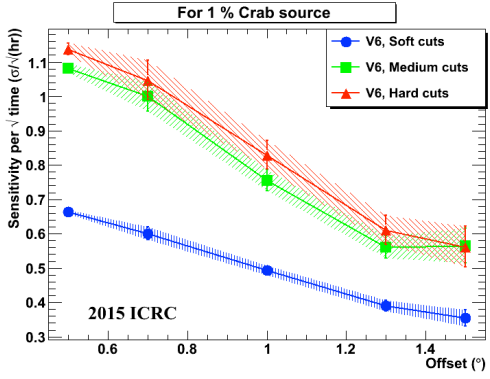 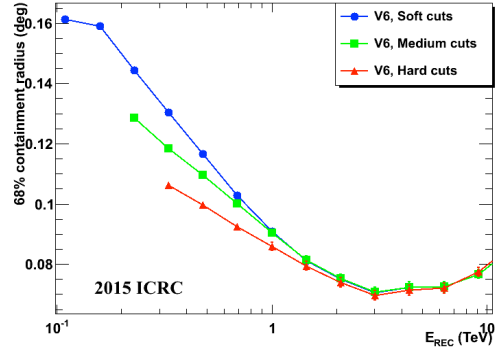

Figure 3: Sensitivity per square-root-hour as a function of wobble offset. Crab Nebula data taken in different offsets with elevation higher than 60° were used to calculate the sensitivity for a weak source. The cuts have different energy threshold values as shown in Figure 2a.

Figure 4: The 68% containment radius in different energy bins. Gamma-ray simulation data for 70° elevation were used to estimate the containment radius for the V6 epoch configuration with different cuts. Events above energy threshold are selected for the plot.

Standard VERITAS observations are carried out with the wobble mode method, in which the source is offset by a small angular distance from the center of the field of view, alternating runs in the four cardinal directions on the sky. For a point source observation, a 0.5° offset is used. However, observations with larger offsets are necessary for studies of extended sources. While sky coverage increases as observing offset increases, the sensitivity and point spread function (PSF) degrade for observations with large offsets. Studying how the performance of the array changes with offset is essential to developing an optimal observing strategy, particularly when dealing with extended regions of gamma-ray emission or multiple sources in the same field of view. Figure 3 shows how the sensitivity of the array in the V6 epoch changes with different offsets for weak sources (1% of Crab Nebula strength) with different sets of cuts by using data taken on the Crab Nebula.

### 3.2 Angular resolution

The angular resolution for a point source at different energies is shown in Figure 4 using simulated events by calculating a radius containing 68% of the gamma-ray events for each bin. The quality of the angular resolution also depends on the azimuthal and zenith angles of the observation. Generally, VERITAS has a 68% containment radius smaller than 0.1° at 1 TeV. Further improvements of angular resolution are possible via tighter event selection cuts, tighter telescope multiplicity requirements or several advanced analysis methods.

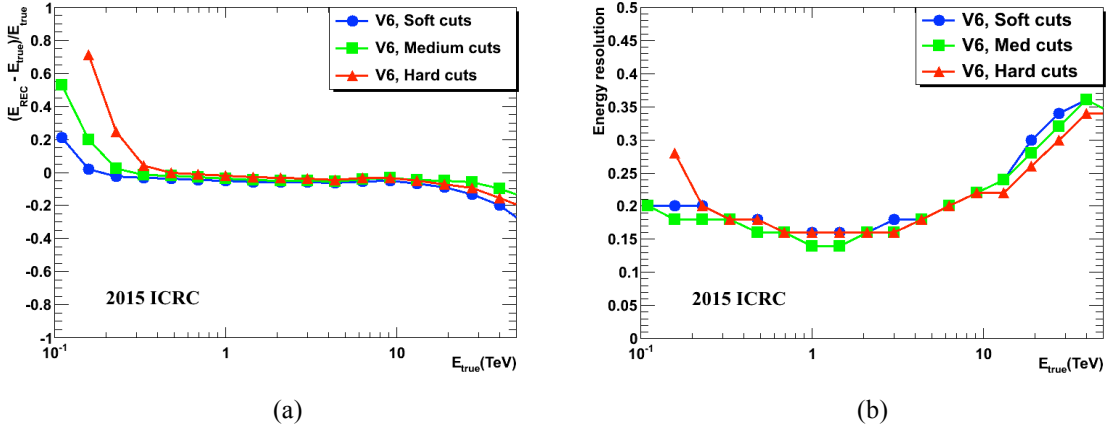

Figure 5: *(a):* Median values of the energy bias $(E_{rec} - E_{true})/E_{true}$ distribution as a function of true energy for the V6 epoch with different sets of cuts. Gamma-ray simulation with elevation 70° was used. *(b):* Energy resolution dependence on true energy for the V6 epoch with different sets of cuts. Energy resolution is defined as the 68% containment radius around the median value of the $(E_{rec} - E_{true})/E_{true}$ distribution. Gamma-ray simulation with elevation 70° was used.

### 3.3 Energy bias and energy resolution

Standard analyses for VERITAS use look-up tables derived from simulated events to reconstruct the energy of incident air showers based on the strength of the signal and the distance of the air shower from the telescope. The difference between true energy and reconstructed energy of a gamma-ray event is studied to estimate the accuracy of the energy reconstruction. Figure 5a shows the bias of the energy reconstruction, which is defined as the median value of the ratio of the difference between true energy and reconstructed energy to the true energy $((E_{rec} - E_{true})/E_{true})$ for standard cuts for the V6 epoch. The energy bias is close to zero for events with energies roughly between the energy threshold value for each set of cuts and a few tens of TeV. The 68% containment width around the median value of the energy bias distribution is defined as the energy resolution. As shown in Figure 5b, the energy resolution of VERITAS is about 15% - 20%. As energy goes higher, the air shower images become more elongated, making it difficult to confine them completely in the camera. Due to this leakage, the energy resolution gets worse as energy goes higher, and increases up to 25% at 10 TeV.

### 4. Performance with advanced analysis methods

The sensitivity of the VERITAS array, particularly to weak sources, can be improved with better event reconstruction methods and with more sophisticated background rejection methods. Among several advanced analysis methods under development for VERITAS (e.g. template-based analysis [13]), we will show the results from a boosted decision tree (BDT) method [14] as an example. A BDT analysis was developed for use on VERITAS data, and it results in a significant improvement in sensitivity. The boosted decision trees are trained on simulated gamma rays and real background events as a function of energy and elevation, and they incorporate the standard

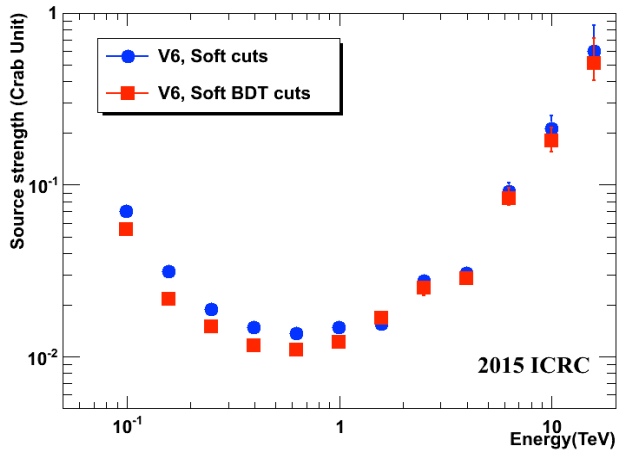

Figure 6: Differential sensitivity comparisons between standard box cuts and BDT cuts for 50 hours of observing time. Crab data taken at high elevation and analyzed with cuts optimized for soft index sources were used for comparisons. Both box cuts and BDT cuts are defined and calculated by using the second analysis package. The same conditions described in Section 3 were used to estimate the differential sensitivity.

Hillas parameters. As shown in the Figure 6, while the improvement of the sensitivity with the BDT method is the most significant at low energy range, the improvements can be seen in the overall energy range. A 10-25% decrease in the observation time required for detection is demonstrated for weak sources with a Crab Nebula-like spectrum and 1% of the Crab Nebula flux. Although the BDT method performs best compared to box cuts on soft spectrum sources, improvements are also observed for sources with hard spectra.

## 5. Summary

A brief summary of the performance of the VERITAS experiment over seven years of operation was presented in this paper. The performance of VERITAS has been improved with multiple upgrades. Further improvements to the sensitivity and performance are expected in the near future as several advanced analysis methods become available.

## 6. Acknowledgements

This research is supported by grants from the U.S. Department of Energy Office of Science, the U.S. National Science Foundation and the Smithsonian Institution, and by NSERC in Canada. We acknowledge the excellent work of the technical support staff at the Fred Lawrence Whipple Observatory and at the collaborating institutions in the construction and operation of the instrument. The VERITAS Collaboration is grateful to Trevor Weekes for his seminal contributions and leadership in the field of VHE gamma-ray astrophysics, which made this study possible.